\documentclass[amsmath,jcp,twocolumn]{revtex4-1}
\usepackage{color}
\usepackage{amsmath}
\usepackage{amsfonts}
\usepackage{amssymb}
\usepackage{graphicx}

\begin{document}
\title{
Exceptional points treatment of cavity spectroscopies
}
\author{Shaul Mukamel}
%\email{smukamel@uci.edu}
%\phone{+1 949 824 7600}
\affiliation{Department of Chemistry, University of California Irvine, Irvine, CA 92697, USA}
\author{Anqi Li}
\affiliation{Department of Chemistry \& Biochemistry, University of California San Diego, La Jolla, CA 92093, USA}
\author{Michael Galperin}
\email{migalperin@ucsd.edu}
%\phone{+1 858 246 0511}
\affiliation{Department of Chemistry \& Biochemistry, University of California San Diego, La Jolla, CA 92093, USA}
%%%%%%%%%%%%%%%%%%%%%%%%%%%%%%%%%%%%%%%%%%
\begin{abstract}
The infrared response of a system of two vibrational modes in a cavity is calculated by an effective 
non-Hermitian Hamiltonian derived by employing the nonequilibrium Green's functions (NEGF) formalism.
Degeneracies of the Hamiltonian (exceptional points, EP) widely 
employed in theoretical analysis of optical cavity spectroscopies are used
in an approximate treatment and compared with the full NEGF.  
Qualitative limitations of the EP treatment are explained by examining
the approximations employed in the calculation. 
\end{abstract}

\maketitle
%%%%%%%%%%%%%%%%%%%%%%%%%%%%%%%%%%%%%%%%%%
%%%%%%%%%%%%%%%%%%%%%%%%%%%%%%%%%%%%%%%%%%%
%%%%%%%%%%%%%%%%%%%%%%%%%%%%%%%%%%%%%%%%%%%

\section{Introduction}
Developments in experimental techniques allow the spectroscopic measurements 
of molecular systems placed in  cavities~\cite{schwartz_reversible_2011,hutchison_modifying_2012,hutchison_tuning_2013,schwartz_polariton_2013,eddins_collective_2014,eizner_temporal_2017,rozenman_long-range_2018,akulov_long-distance_2018,xiang_intermolecular_2020,chen_cavity-enabled_2022,yang_molecular_2022}.
Cavity confinement promotes strong coupling between radiation field and molecular 
degrees of freedom. The optical control of molecular
responses (conformational switching, charge and energy transfer, 
chemical reactions rates) thus becomes feasible.
To enhance the signal, the majority of
such experiments are performed on samples containing many molecules.
However, spectroscopic measurements of single-molecules in 
plasmonic nanocavities were reported as well~\cite{chikkaraddy_single-molecule_2016,benz_single-molecule_2016}.

Theoretical considerations of spectroscopy of molecules placed in optical cavities 
usually rely on simplified model approaches. 
While the actual experimental setup deals with open driven quantum mechanical
systems, the simplest theoretical treatments use Hermitian Hamiltonian for the cavity mode 
and molecular degrees of freedom~\cite{f_ribeiro_theory_2018}. 
The open character of the system is introduced later in an ad hoc manner 
either using single particle scattering theory or within
the input-output methodology, 
to name the two most popular and closely related approaches~\cite{lehmann_superposition_1996,ciuti_input-output_2006}.
In such treatments one diagonalizes the Hamiltonian matrix represented in a basis
of isolated molecule many-body states with direct product with harmonic oscillator states representing
the cavity mode. Its eigenstates mix different degrees of freedom of the system. 
The mixture of light and matter degrees of freedom are called polaritons. 

More advanced treatments take into account the open character of the system 
already at the level of the Hamiltonian formulation - an 
effective non-Hermitian Hamiltonian consideration. 
Non-Hermitian quantum mechanics  is implemented in many research areas from optics, 
to quantum field theory, to molecular physics~\cite{moiseyev_non-hermitian_2011,rodriguez_classical_2016}.
Non-Hermitian quantum mechanics is obtained by adding
complex absorbing potentials (CAPs) to Hermitian Hamiltonians. 
CAPs which represent the open character of the system
are often employed to simplify numerical simulations. 
System responses are especially non-trivial at the degeneracy points
of the non-Hermitian Hamiltonian spectrum known
as exceptional points (EPs)~\cite{berry_physics_2004,gunther_projective_2007,heiss_physics_2012}.
The concept is widely applied in the description of open quantum 
systems~\cite{rotter_non-hermitian_2009,garmon_analysis_2012}
including studies of quantum transport at junctions~\cite{toroker_relation_2009,rotter_review_2015}.
It is also popular in optics and polaritonics~\cite{delga_theory_2014,gao_observation_2015,miri_exceptional_2019,ergoktas_topological_2022,soleymani_chiral_2022,finkelstein-shapiro_non-hermitian_2022}
where EPs where shown to be responsible for exotic phenomena such as 
decreasing intensity of the emitted laser light for increasing pump power~\cite{brandstetter_reversing_2014},
unidirectional transport~\cite{guo_observation_2009},
chiral modes~\cite{dembowski_observation_2003,peng_chiral_2016}, and
anomalous lasing~\cite{sun_experimental_2014,peng_loss-induced_2014}.
In addition, topological structures~\cite{dembowski_experimental_2001,lee_observation_2009,choi_quasieigenstate_2010,xu_topological_2016}, phase transitions~\cite{jung_phase_1999,eleuch_open_2014},
and topological Berry phase~\cite{heiss_phases_1999,dembowski_encircling_2004} 
are observed in the vicinity of exceptional points.

First principles treatment of open quantum systems starts with a Hermitian quantum
mechanical description of the universe. 
Description of the open system is obtained by tracing out environmental (bath) degrees of freedom. 
The nonequilibrium Green's function (NEGF)
approach to open quantum systems follows this paradigm~\cite{haug_quantum_2008,stefanucci_nonequilibrium_2013}.
The influence of environment is then incorporated through the self-energies.

Here, we study a model of two vibrational modes in a cavity. 
The vibrations are not directly coupled but are coupled 
to the cavity mode and to thermal (phonon) baths.
The cavity mode is driven by laser modeled as a continuum of radiation modes
narrowly populated around the laser frequency.
The model was discussed previously
within the non-Hermitian quantum mechanics paradigm where
effective non-Hermitian Hamiltonian derived within the input-output formalism 
consideration was used for consideration of the role
of exceptional points in responses of the open quantum system~\cite{yang_phonon_2020}.
Here we adopt a non-Hermitian quantum mechanics formulation
starting with the exact NEGF description.  
We then discuss concept of exceptional points 
and compare the predictions of the non-Hermitian simulations 
to those of the more rigorous full NEGF results.

The structure of the paper is as follows. In Section~\ref{model}
we introduce the model, present corresponding NEGF formulation, 
and use it to derive non-Hermitian quantum mechanical description.
Section~\ref{numres} presents numerical simulations performed
within NEGF and within exceptional points approaches 
and discusses similarities and differences of the two methods.  
Conclusions are drawn in Section~\ref{conclude}.

%%%%%%%%%%%%%%%%%%%%%%%%%%%%%%%%%%%%%%%%%%%
%%%%%%%%%%%%%%%%%%%%%%%%%%
%%%%% QM description of laser field %%%%%%
%%%%%%%%%%%%%%%%%%%%%%%%%%
\section{Two vibrational modes in a cavity}\label{model}
\subsection{Model}
\begin{figure}[htbp]
\centering\includegraphics[width=\linewidth]{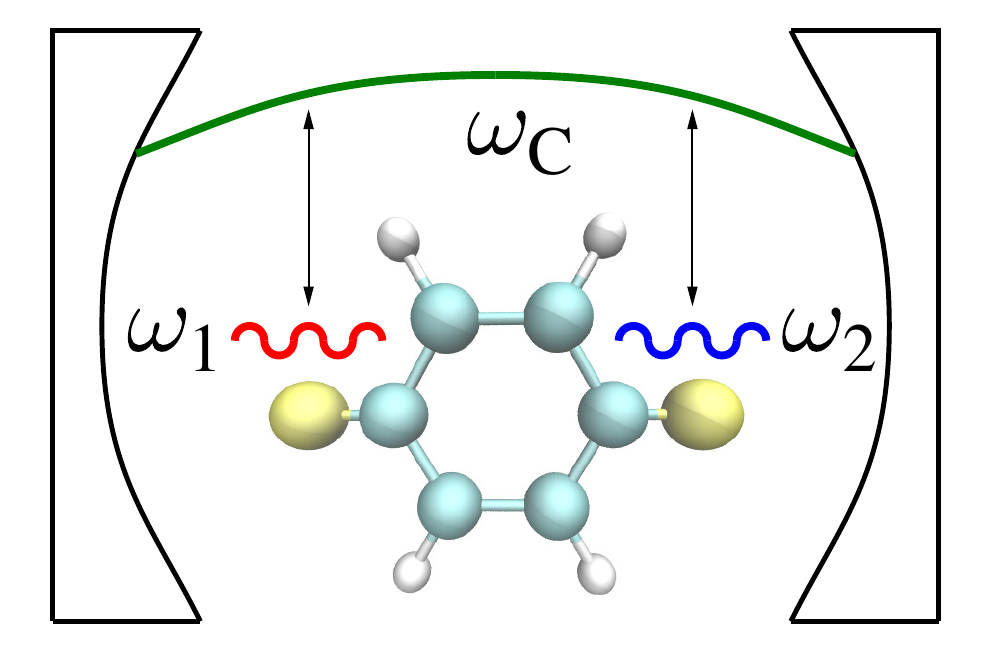}
\caption{\label{fig1}
 Sketch of a single molecule in a cavity model.
}
\end{figure}

We consider two vibrational modes $\omega_1$ and $\omega_2$ 
coupled to the same cavity mode $\omega_C$ and driven by 
the same laser field (see Fig.~\ref{fig1}). 
The vibrational modes are coupled further to thermal baths.
In the rotating wave approximation the Hamiltonian of the system is 
\begin{equation}
\hat H=\hat H_0 + \hat V
\end{equation}
where 
\begin{equation}
\begin{split}
\hat  H_0 &= \omega_C\hat a_C^\dagger\hat a_C
+ \sum_{i=1,2}\omega_i\hat b_i^\dagger\hat b_i 
\\ &
+ \sum_{\alpha}\omega_\alpha \hat a^\dagger_\alpha \hat a_\alpha 
+ \sum_{i=1,2}\sum_{\beta_i} \omega_{\beta_i} \hat b^\dagger_{\beta_i} \hat b_{\beta_i}
\\
\hat V &= 
\sum_i g_i\hat a_C^\dagger\hat a_C\left(\hat b_i+\hat b_i^\dagger\right)
\\ &
+\sum_\alpha\left(V_{C\alpha} \hat a_C^\dagger\hat a_\alpha
+ V_{\alpha C}\hat a_\alpha^\dagger\hat a_C\right)
\\ &
+\sum_{i=1,2}\sum_{\beta_i}\left(V_{i\beta_i}\hat b_i^\dagger\hat b_{\beta_i}
+ V_{\beta_i i}\hat b_{\beta_i}^\dagger\hat b_i\right)  
\end{split}
\end{equation}
Here, the boson operators $\hat a^\dagger_C$ and $\hat b^\dagger_i$ 
create quanta of cavity mode excitation and vibration $i$, respectively.
$\hat a^\dagger_\alpha$ and $\hat b^\dagger_{\beta_i}$
describe excitations of radiation field and phonon (thermal) baths, respectively.

\subsection{The NEGF formulation}
We first consider the cavity mode and vibrations by ignoring mixing between 
the cavity mode and vibrational degrees of freedom. 
The corresponding single-particle Green's functions are defined on the Keldysh contour as
\begin{equation}
\label{GF_def}
\begin{split}
A_C(\tau_1,\tau_2) &\equiv -i\left\langle T_c\, \hat a_C(\tau_1)\,\hat a_C^\dagger(\tau_2)\right\rangle
\\
B_{ij}(\tau_1,\tau_2) &\equiv -i\left\langle T_c\, \hat b_i(\tau_1)\,\hat b_j^\dagger(\tau_2)\right\rangle\qquad (i,j=1,2)
\end{split}
\end{equation}
Here, $\tau_{1,2}$ are contour variables and $T_c$ is the contour ordering operator.
These Green's functions satisfy the Dyson equations
\begin{equation}
\label{GF_Dyson}
\begin{split}
&\left(i\frac{\partial}{\partial\tau_1}-\omega_C\right) A_C(\tau_1,\tau_2)
= \delta(\tau_1,\tau_2) 
\\ &\qquad\qquad\qquad
+ \int_c d\tau\, \Sigma_A(\tau_1,\tau)\, A_C(\tau,\tau_2)
\\
&\left(i\frac{\partial}{\partial\tau_1}\mathbf{I}-\mathbf{\Omega}\right) \mathbf{B}(\tau_1,\tau_2)
= \delta(\tau_1,\tau_2)\,\mathbf{I} 
\\ &\qquad\qquad\qquad
+ \int_c d\tau\, \mathbf{\Sigma}_B(\tau_1,\tau)\, \mathbf{B}(\tau,\tau_2)
\end{split}
\end{equation}
where $\mathbf{I}$ is unity matrix,
\begin{equation}
\mathbf{\Omega}=\begin{pmatrix}
\omega_1 & 0 \\ 0 & \omega_2
\end{pmatrix},
\end{equation}
and 
\begin{equation}
\label{SigmaAB}
\begin{split}
\Sigma_{A}(\tau_1,\tau_2) &= \Sigma_{rad}(\tau_1,\tau_2) + \Sigma_{vib}(\tau_1,\tau_2)
\\
\mathbf{\Sigma}_{B}(\tau_1,\tau_2) &= \mathbf{\Sigma}_{ph}(\tau_1,\tau_2) 
+ \mathbf{\Sigma}_{C}(\tau_1,\tau_2)
\end{split}
\end{equation}
are self-energies. The latter account for effect of
couplings to the radiation field and phonon baths, 
\begin{equation}
\label{Sigma_rad_ph}
\begin{split}
\Sigma_{rad}(\tau_1,\tau_2) &= \sum_{\alpha} V_{C\alpha}\, A_\alpha(\tau_1,\tau_2)\, V_{\alpha C}
\\
\left[\Sigma_{ph}\right]_{ii}(\tau_1,\tau_2) &= \sum_{\beta_i} V_{i\beta_i}\, B_{\beta_i}(\tau_1,\tau_2)\,
V_{\beta_i i}
\end{split}
\end{equation}
and for interaction between the cavity mode and vibrational degrees of freedom.
In the second order diagrammatic expansion the latter are
\begin{equation}
\label{Sigma_vib_c}
\begin{split}
\Sigma_{vib}(\tau_1,\tau_2) &= i\sum_{i,j}g_i\bigg(
B_{ij}(\tau_1,\tau_2)+B_{ij}(\tau_2,\tau_1)
\bigg)g_j
\\
\left[\Sigma_{C}\right]_{ij}(\tau_1,\tau_2) &=i\, g_i\, P(\tau_1,\tau_2)\, g_j
\end{split}
\end{equation}
 $A_\alpha(\tau_1,\tau_2)$ and $B_{\beta_i}(\tau_1,\tau_2)$
are respectively the Green's function of free photons and phonons, 
\begin{equation}
\begin{split}
A_\alpha(\tau_1,\tau_2) &\equiv -i\left\langle T_c\, \hat a_\alpha(\tau_1)\,
\hat a_\alpha^\dagger(\tau_2) \right\rangle_0
\\
B_{\beta_i}(\tau_1,\tau_2) &\equiv -i\left\langle T_c\, \hat b_{\beta_i}(\tau_1)\,
\hat b_{\beta_i}^\dagger(\tau_2) \right\rangle_0,
\end{split}
\end{equation}
and
\begin{equation}
\label{bubble}
P(\tau_1,\tau_2) \equiv - \langle T_c\, \hat a_C^\dagger(\tau_1)\hat a_C(\tau_1)\,
 \hat a_C^\dagger(\tau_2)\hat a_C(\tau_2)\rangle
\end{equation}
is the polarization bubble (two-particle Green's function of the cavity mode
excitations). 

To compute the Green's functions (\ref{GF_def}) one has to solve 
the system of coupled Dyson equations
(\ref{GF_Dyson}) self-consistently on the Keldysh contour.
Because the self-energy $\Sigma_c$ in Eq.~(\ref{Sigma_vib_c}) 
is expressed in terms of two-particle Green's function
(\ref{bubble}) simultaneous solution of the Bethe-Salpeter equation should be considered for the polarization
bubble. To simplify the analysis we neglect effect of vibrational degrees of freedom
on cavity mode. In this case the self-energy $\Sigma_{vib}(\tau_1,\tau_2)$ 
in Eq.(\ref{Sigma_vib_c})
can be dropped and the polarization bubble (\ref{bubble}) can be approximately expressed in terms of
single-particle Green's functions $A_C$ using the Wick's theorem
\begin{equation}
\label{SigmaC}
\left[\Sigma_C\right]_{ij}(\tau_1,\tau_2) \approx i\, g_i\, A_C(\tau_1,\tau_2)\, A_C(\tau_2,\tau_1)\, g_j
\end{equation}
%%%%%%%%%%%%%%%%%%%%%%%%%%
\subsection{Linearization of the polarization bubble}
After neglecting back action of vibrational degrees of freedom on cavity mode, 
assuming wide band approximation for coupling to radiation field modes, considering steady-state,
and taking Fourier transform we get from (\ref{SigmaAB}) for retarded and lesser projections of $\Sigma_A$
\begin{equation}
\begin{split}
\Sigma_A^{r}(E) &=-\frac{i}{2}\kappa
\\
\Sigma_A^{<}(E) &=-i\kappa\,\frac{P}{\omega_L}\,\frac{\delta}{(E-\omega_L)^2+(\delta/2)^2}
\end{split}
\end{equation}
Here, $\kappa$ is cavity mode dissipation, $\omega_L$ is the laser frequency, $\delta$ is the laser linewidth,
and $P$ is the laser intensity. Thus,
\begin{equation}
\begin{split}
\langle \hat a_C^\dagger\hat a_C\rangle &= 
i\, A_C^{<}(t,t)
\\ &
=i\int\frac{dE}{2\pi}\, A_C^r(E)\,\Sigma_A^{<}(E)\, A_C^a(E)
\\ &\overset{\delta\to0+}{\longrightarrow} \kappa\, \frac{P}{\omega_L}\, A_C^r(\omega_L)\, A_C^a(\omega_L)
\end{split}
\end{equation}

We split the total cavity field into the pumped field $\alpha_c$ (treated classically) and
and fluctuations $\delta \hat a_c$ due to presence of empty radiation background (treated quantum mechanically)
\begin{equation}
\label{aC_split}
 \hat a_C = \alpha_C + \delta\hat a_C
\end{equation}
Thus,
\begin{equation}
\label{alpha_c}
\begin{split}
& \langle\hat a_C^\dagger\hat a_C\rangle =  \alpha_C^{*}\,\alpha_C\quad\Rightarrow
 \\ 
& \alpha_C = \sqrt{\kappa\frac{P}{\omega_L}}\,\frac{1}{\omega_L-\omega_C+\frac{i}{2}\kappa}
 \equiv \epsilon\,\frac{1}{\Delta+\frac{i}{2}\kappa}
\end{split}
\end{equation}
where $\epsilon\equiv\sqrt{\kappa P/\omega_L}$ is the driving strength and 
$\Delta\equiv\omega_L-\omega_C$ is the detuning between the pumping laser field and cavity mode frequency.

Linearization of the polarization bubble (\ref{bubble}) yields
\begin{equation}
\label{bubble_lin}
\begin{split}
& P(\tau_1,\tau_2) 
\\
&= - \left\langle T_c\,
 \left(\alpha_C^{*}+\delta a_C^\dagger(\tau_1)\right)\left(\alpha_C+\delta\hat a_C(\tau_1)\right)\right.
 \\ &\qquad\,\,\,\left.\times
 \left(\alpha_C^{*}+\delta a_C^\dagger(\tau_2)\right)\left(\alpha_C+\delta a_C(\tau_2)\right)\right\rangle
 \\ 
 &\approx -i\,\alpha_C^{*}\,\alpha_C\bigg(\mathcal{A}_C(\tau_1,\tau_2)+\mathcal{A}_C(\tau_2,\tau_1)\bigg)
\end{split}
\end{equation}
where
\begin{equation}
\label{calAc}
\mathcal{A}_C(\tau,\tau')\equiv -i\langle T_c\,\delta\hat a_C(\tau)\,\delta\hat a_C^\dagger(\tau')\rangle
\end{equation}
is the single-particle Green's function of cavity mode fluctuations coupled to an empty continuum of 
radiation field modes. Green's function (\ref{calAc}) satisfies the Dyson equation
\begin{equation}
\label{calAc_Dyson}
\begin{split}
\bigg(i\frac{\partial}{\partial\tau_1}+\Delta\bigg)&\mathcal{A}_C(\tau_1,\tau_2) =
\delta(\tau_1,\tau_2) 
\\ &
+\int_c d\tau\, \Sigma_{rad}^{empty}(\tau_1,\tau)\,\mathcal{A}_C(\tau,\tau_2)
\end{split}
\end{equation}
\iffalse
Note that such linearization procedure besides dropping higher order correlations 
also misses inter-relation between two parts of the cavity field. The latter may become important 
at low laser intensities where cavity field quantum features become pronounced.
\fi
Separation of the field into two parts with one, $\alpha_C$, 
treated classically and the other, $\delta\hat\alpha_C$, quantum mechanically 
introduces several approximations:
\begin{enumerate}
\item The main difference between classical and quantum fields is the ability
of the latter to mediate photon supported effective interaction between
quantum degrees of freedom in the system. For the model considered here,
quantum degrees of freedom interacting via photon are molecular vibrations,
and photon induced interaction is described by $\Sigma_C$.
Expressions for the interaction, Eq.(\ref{SigmaC}) for NEGF and Eq.(\ref{Sigma_c_ss}) below for EP,
become very different in strong fields, where neglected quantum character of 
the $\alpha_C$ part becomes pronounced. 
\item A classical treatment misses all quantum correlations. This is reflected by
time-local character of $\alpha_C$ contribution vs. time-non-local correlation functions
in quantum treatment.
\item $\alpha_C$ and $\delta\hat\alpha_C$ of Eq.(\ref{aC_split}) are two parts representing 
the same radiation field mode. Thus, in the case of spontaneous emission, 
described within the $\delta\hat\alpha_C$ part, resulting photon should 
be accounted for in the $\alpha_C$ part. In the linearized formulation 
such photon is disregarded. The effect should be significant at weak laser fields.
\end{enumerate}

This completes the description of the cavity mode. 
Below we focus on dynamics of the vibrational degrees of freedom. 
The result for the linearized polarization bubble $P$, Eq.(\ref{bubble_lin}), can now be used in expression for self-energy of vibrations due to coupling to the cavity mode
$\Sigma_C$, Eq.(\ref{Sigma_vib_c})
%%%%%%%%%%%%%%%%
\subsection{Exceptional points}
To introduce the concept of exceptional points one has to formulate 
the dynamics of the two vibrational degrees of freedom in terms of some 
effective non-Hermitian Hamiltonian 
$\hat H_{eff}$.
Within the NEGF (assuming steady-state) 
this is equivalent to substituting proper on-the-contour Dyson equation
for the Green's function $\mathbf{B}$,
\begin{equation}
\label{B_Dyson}
\begin{split}
&\begin{bmatrix}
E\,\mathbf{I}-\mathbf{\Omega} - \mathbf{\Sigma}_B^{c}(E) & \mathbf{\Sigma}_B^{<}(E) \\
\mathbf{\Sigma}_B^{>}(E) & -E\,\mathbf{I}+\mathbf{\Omega}-\mathbf{\Sigma}_B^{\tilde c}(E)
\end{bmatrix}
\\ &\qquad\qquad\qquad\quad\times
\begin{bmatrix}
\mathbf{B}^{c}(E) & \mathbf{B}^{<}(E) \\
\mathbf{B}^{>}(E) & \mathbf{B}^{\tilde c}(E)
\end{bmatrix} = \mathbf{I},
\end{split}
\end{equation}
with an effective equation-of-motion,
\begin{equation}
\label{B_Heff}
\begin{bmatrix}
E\,\mathbf{I}-\mathbf{H}_{eff}  & 0 \\
0 & -E\,\mathbf{I}+\mathbf{H}_{eff}
\end{bmatrix}
\begin{bmatrix}
\mathbf{B}^{c}(E) & \mathbf{B}^{<}(E) \\
\mathbf{B}^{>}(E) & \mathbf{B}^{\tilde c}(E)
\end{bmatrix} = \mathbf{I},
\end{equation}
Here, $c$, $<$, $>$, and $\tilde c$ indicate causal, lesser, greater, and anti-causal projections, respectively.
Taking into account that
\begin{equation}
\begin{split}
\mathbf{\Sigma}_B^{c}(E) &
%\equiv \theta(t_1-t_2) \mathbf{\Sigma}^{B\, >}(t_1,t_2)+\theta(t_2-t_1)\mathbf{\Sigma}^{B\, <}(t_1,t_2) 
= \mathbf{\Sigma}_B^{r}(E) + \mathbf{\Sigma}_B^{<}(E)
\\
\mathbf{\Sigma}_B^{\tilde c}(E) &
%\equiv \theta(t_1-t_2) \mathbf{\Sigma}^{B\, <}(t_1,t_2)+\theta(t_2-t_1)\mathbf{\Sigma}^{B\, >}(t_1,t_2) 
= \mathbf{\Sigma}_B^{a}(E) + \mathbf{\Sigma}_B^{>}(E)
\end{split}
\end{equation}
 (here $r$ and $a$ are retarded and advanced projections)
transition from (\ref{B_Dyson}) to (\ref{B_Heff}) is only meaningful 
under the following assumptions:
\begin{enumerate}
\item\label{cond1} 
Lesser projection of the self-energy is disregarded
\[
\mathbf{\Sigma}_B^{<}(E)\equiv\mathbf{\Sigma}_{ph}^{<}(E)+\mathbf{\Sigma}_C^{<}(E)=0
\]
\begin{enumerate}
\item Thermal (phonon) bath contribution can be disregarded 
when the bath is held at zero temperature,  that is $\mathbf{\Sigma}_{ph}^{<}(E)=0$. 
\item Neglecting contribution $\mathbf{\Sigma}_C^{<}(E)$ is consistent
with the assumption of the field splitting, so that $\langle\delta\hat a_C^\dagger\,\delta\hat a_C\rangle=0$, but requires also assuming 
$\langle\delta\hat a_C\,\delta\hat a_C^\dagger\rangle=0$
(i.e. neglect quantum character of the radiation background)
which is hard to justify.
\end{enumerate}
\item\label{cond2} Greater projection of the self-energy is disregarded
\[
\mathbf{\Sigma}_B^{>}(E)\equiv\mathbf{\Sigma}_{ph}^{>}(E)+\mathbf{\Sigma}_C^{>}(E)=0
\]
This assumption is equivalent to neglect of  quantum effects in the bath
which contradicts zero temperature requirement of the previous step.
Indeed, considering lesser and greater 
projections of phonon self-energy $\Sigma_{ph}$: 
$\Sigma_{ph}^{<}(E)=-i \gamma(E) N(E)$ and 
$\Sigma_{ph}^{>}(E)=-i \gamma(E) [1+N(E)]$
(here, $\gamma(E)$ the dissipation rate and $N(E)$ is the Bose-Einstein 
phonon distribution in thermal bath) one sees that when $T\to 0\,\Rightarrow\, N(E)\to 0$. 
This justifies neglect of the lesser projection. However, quantum effects
in principle do not allow similar neglect of the greater projection 
of the self-energy: $\Sigma_{ph}^{E}\overset{T\to 0}{\rightarrow} -i\gamma(E)$.
\item\label{cond3} On the scale of the system relevant energies energy dependence of the self-energy projections is smooth, so that
\[
\mathbf{\Sigma}_B^{r/a}(E)\approx \mathbf{\Sigma}_B^{r/a}(E_0)=const
\]
for some arbitrary energy $E_0$.
\end{enumerate}

To obtain an effective vibrational Hamiltonian we only need to know
the retarded projection of the self-energy $\mathbf{\Sigma}_B$, which
according to Eq.(\ref{SigmaAB}) has contributions from coupling to thermal bath, 
$\mathbf{\Sigma}_{ph}^{r}$, and interaction via cavity mode,
$\mathbf{\Sigma}_C^{r}$. Neglecting bath induced correlations and assuming 
wide band approximation (WBA) the former is
\begin{equation}
\label{Sigma_ph_ss}
 \mathbf{\Sigma}_{ph}^{r}(E) = -\frac{i}{2}
 \begin{bmatrix}
 \gamma_1 & 0 \\ 0 & \gamma_2
 \end{bmatrix}
\end{equation}
Here $\gamma_i(E)\equiv2\pi\sum_{\beta_i} V_{i\beta_i}\, V_{\beta_i i}\,\delta(E-\omega_{\beta_i})$ ($i=1,2$)
is the dissipation rate of vibrational mode $i$.
The latter is obtained using (\ref{bubble_lin}) and (\ref{calAc_Dyson}) 
in (\ref{Sigma_vib_c})
\begin{align}
\label{Sigma_c_ss}
 \Sigma_C^{r}(E) &= g_i\, \alpha_C^{*}\,\alpha_C\, g_j \bigg(\mathcal{A}_C^r(E)+\mathcal{A}_C^a(-E)\bigg)
 \\ &= g_i\, \alpha_C^{*}\,\alpha_C\, g_j \bigg(
 \frac{1}{\Delta+E+\frac{i}{2}\kappa} + \frac{1}{\Delta -E-\frac{i}{2}\kappa}
 \bigg)
\nonumber
\end{align}

Finally, utilizing (\ref{alpha_c}), taking $g_1=g_2\equiv g$, choosing 
an arbitrary value $E_0$ in  (\ref{Sigma_c_ss}), 
and adding (\ref{Sigma_ph_ss}) yields effective Hamiltonian 
\begin{equation}
\label{Heff}
\mathbf{H}_{eff} = \begin{bmatrix}
\omega_1 - \frac{i}{2}\gamma_1 + \Lambda & \Lambda \\
\Lambda & \omega_2 - \frac{i}{2}\gamma_2 + \Lambda
\end{bmatrix}
\end{equation}
where
\begin{equation}
\begin{split}
\Lambda &\equiv g^2\, \frac{P}{\omega_L}\,\frac{\kappa}{\Delta^2+\frac{\kappa^2}{4}}
\\ & \times
\left[\frac{\Delta+E_0-\frac{i}{2}\kappa}{\left(\Delta+E_0\right)^2+\frac{\kappa^2}{4}}
+\frac{\Delta-E_0+\frac{i}{2}\kappa}{\left(\Delta-E_0\right)^2+\frac{\kappa^2}{4}}
\right]
\end{split}
\end{equation}
The eigenenergies of the Hamiltonian are 
\begin{equation}
\begin{split}
E_{\pm}&=\frac{\omega_1-\frac{i}{2}\gamma_1+\Lambda+\omega_2-\frac{i}{2}\gamma_2+\Lambda}{2}
\\ &
\pm\frac{1}{2}\sqrt{\left[\left(\omega_1-\frac{i}{2}\gamma_1\right)-\left(\omega_2-\frac{i}{2}\gamma_2\right)\right]^2+4\Lambda^2}
\end{split}
\end{equation}
Thus, exceptional point (defined as point of degeneracy for eigenvalues) is
\begin{equation}
\label{EP}
\left[\left(\omega_1-\frac{i}{2}\gamma_1\right)-\left(\omega_2-\frac{i}{2}\gamma_2\right)\right]^2+4\Lambda^2=0
\end{equation}
These results were first obtained in Ref.~\cite{yang_phonon_2020} 
using the input-output formalism.

To re-introduce information on the baths, the input-output formalism 
utilizes concept of noise operators. Their correlation functions
yield lesser and greater projections of the corresponding self-energies,
$\Sigma_{rad}^{\gtrless}(E)$ and $\Sigma_{ph}^{\gtrless}(E)$,
usually taken at a particular fixed energy $E_0$ (delta-correlated in time).
In this respect, the input-output formalism is similar to the single-particle scattering theory.
Note that lesser and greater projections of the self-energy due to 
coupling between vibrational and cavity modes,  $\Sigma_{C}^{\gtrless}(E)$,
are not taken into account by the procedure and thus are disregarded in the EP
treatment. Strictly speaking such consideration is inconsistent,
because retarded projection of the same self-energy, $\Sigma_C^r(E)$,
plays central role in the EP approach. 
The three projections are not independent, 
all three are related to the same self-energy defined on the Keldysh contour.
Thus, keeping one of the projections while ignoring the others is inconsistent.
In terms of physical effects, retarded projection yields information on the level shift 
and dissipation, while lesser and greater projections describe ability to 
in- and out-scattering of energy quanta. Ignoring the latter while keeping the former 
is equivalent to neglect of energy exchange between molecular degrees of freedom.
Also, it leads to violation of the fluctuation-dissipation relations.
Thus, besides disregarding energy transfer between molecular vibrations
due to radiation field-induced interaction, such description will not allow to 
reach correct thermal equilibrium.

Knowledge of the $B$ Green's functions, Eq.(\ref{GF_def}), 
within the NEGF formalism
or effective Hamiltonian $\hat H^{eff}$, Eq.(\ref{Heff}), within the EP formalism 
allows to simulate multiple system characteristics (energy flux, populations of
the modes, spectrum, effective temperatures, etc.). Below, following 
Ref.~\cite{yang_phonon_2020} we focus on simulation of the spectra
of vibrational degrees of freedom and their effective temperatures.

%%%%%%%%%%%%%%%%%%%%%%%%%%%%%%%%%%%%%%%%%%%%%%%%%%%%%

%%%%%%%%%%%%%%%%%%%%%%%%%%%%%%%%%%%%%%%%%%%
\section{Numerical results}\label{numres}
While the EP approach is widely applied in
optomechanics~\cite{aspelmeyer_cavity_2014,xu_topological_2016,bemani_synchronization_2017,sheng_self-organized_2020,yang_phonon_2020}
its applicability in single-molecule cavity systems~\cite{benz_single-molecule_2016,chikkaraddy_single-molecule_2016} is not well established.
Here, we compare calculations obtained utilizing exceptional points (EP)
with those of the full NEGF simulations for
a single-molecule in a cavity at steady-state regime. 

Following Ref.~\cite{yang_phonon_2020} we focus on simulation of the spectra
of vibrational degrees of freedom and of their effective temperatures.
The former is defined by lesser projection of the Green function $B$.
At steady state,
\begin{equation}
\label{spectrum}
 S_i(E)\equiv i\, B_{ii}^{<}(E)\qquad (i=1,2)
\end{equation}
Usually exceptional points indicate thresholds for significant changes
in the system response. In particular, for our model only one EP (\ref{EP}) is possible 
It was shown in Ref.~\cite{yang_phonon_2020}
that this exceptional point yields a critical value of effective 
inter-mode coupling $\Lambda$. Coupling strengths below the critical value
provide spectrum with peaks at the vibrational mode frequencies.
Couplings above the critical value yield splitting in the spectrum.

Effective temperature is obtained from assumption of the thermal vibrational distribution
\begin{equation}
\label{Teff}
 k_BT_i^{eff}\equiv\frac{\hbar\omega_i}{\log\left(1+\left[\int\frac{dE}{2\pi}\, S_i(E)\right]^{-1}\right)}
\end{equation}
In derivation of Eq.(\ref{Teff}) we use the Bose-Einstein
distribution instead of its limiting value $k_BT_i^{eff}/\hbar\omega_i$
used in Ref.~\cite{yang_phonon_2020} which does not apply
for our choice of parameters.

We perform the simulations using typical parameters for 
a single molecule in a cavity setup:
cavity mode with frequency $\omega_C=2$~eV 
and escape rate $\kappa=10^{-3}$~eV is pumped with a monochromatic laser
with frequency $\omega_L=1.5$~eV and linewidth $\delta=10^{-5}$~eV. 
Molecular vibrations $\omega_1=0.1$~eV and $\omega_2=0.08$~eV
are coupled two different thermal baths. Eneregy escape rates to 
the baths are $\gamma_1=10^{-3}$~eV and $\gamma_2=5\times 10^{-4}$~eV.
Coupling strengths of the vibrations to the cavity mode are $g_1=g_2=0.05$~eV.
Realistic laser intensities used in optical experiments with single molecule 
junctions are $\sim 1$~kW/cm${}^2$~\cite{ward_vibrational_2011}.
This together with characteristics cross-section of $3$~nm${}^2$
and laser frequency $\omega_L=1.5$~eV ($\lambda_L\sim 780$~nm) yields
$P/\omega_L\sim 10^4$~eV.

\begin{figure}[htbp]
\centering\includegraphics[width=\linewidth]{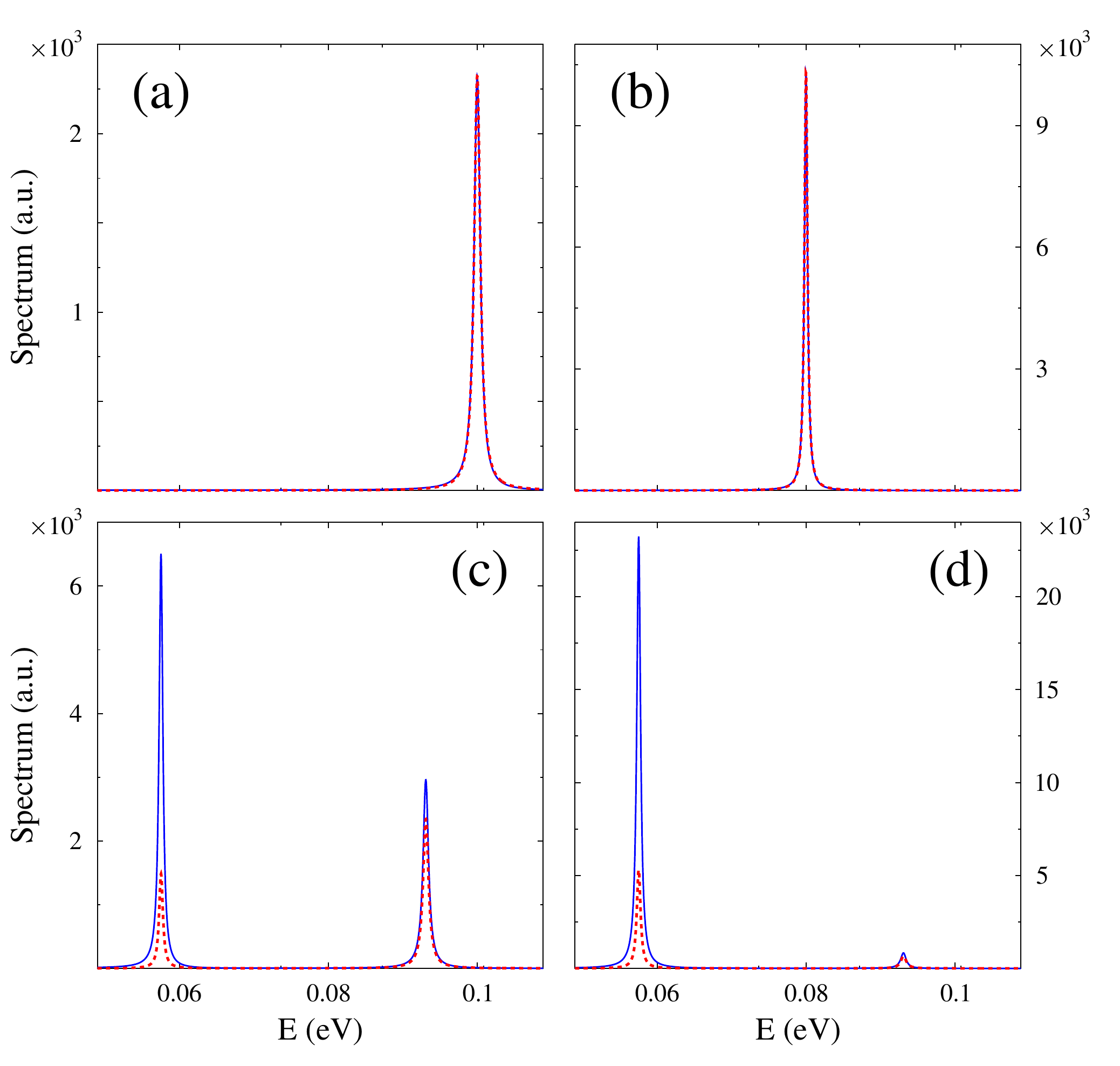}
\caption{\label{fig2}
Vibrational spectrum $S(E)$, Eq.(\ref{spectrum}), in a cavity.
Shown are results of calculations within EP (dashed line, red) and
NEGF (solid line, blue) approaches for $\omega_1$ with inter-mode
coupling (a) below and (c) above exceptional point value.
Similar results for $\omega_2$ are shown in panels (b) and (d), respectively.
See text for parameters.
}
\end{figure}

For the chosen parameters the critical value of inter-mode coupling, Eq.(\ref{EP}), is
$\Lambda\sim 0.01$~eV which correspond to $P/\omega_L\sim 10^2$~eV.
Figure~\ref{fig2} shows the infrared vibrational spectrum, Eq.(\ref{spectrum}),
for effective coupling $\Lambda$ below (top row, $P/\omega_L=10$~eV) and
above (bottom row, $P/\omega_L=10^4$~eV) its critical value.
As expected, the spectrum shows mode splitting for values of the 
coupling above the exceptional point, Eq.(\ref{EP}).
For the chosen parameters EP and NEGF yield similar results for 
$P/\omega_L=10$~eV, the
differences appear at strong couplings: 
results for heights of peaks in the spectrum (including qualitative relative peak values)
are predicted differently by the two approaches.
An obvious reason for the discrepancy is neglect of self-energy
$\Sigma_C^{<}$ in the EP formulation which becomes pronounced
for stronger couplings. 
Disregarding $\Sigma_C^{<}$ leads to the neglect of energy exchange between molecular vibrations 
and violates fluctuation-dissipation relations. 
One sees that for the particular choice of parameters discrepancy in
prediction of position and width of peaks is relatively small.
This is expected because retarded projection of self-energy $\Sigma_B$, Eq.(\ref{SigmaAB}),
to which $\Sigma_C^r$ contributes is responsible for peaks shifts (real part of $\Sigma_B^r$)
and widths (imaginary part of $\Sigma_B^r$).
Note however that by its very construction the EP approach disregards 
self-consistency of the complete NEGF treatment (see discussion in the conclusions).
This lack of self-consistency results in incorrect $\Sigma_C^r$ contribution.
Note also that because the EP misses quantum correlations between vibrations
(consequence of the linearization procedure),
another choice of parameters may lead to different results for
the two approaches also for small values of inter-mode coupling. 

While low power results coincide for the parameters chosen, we note that this cannot be considered
as validation of the EP approach. Close or coinciding results at a particular 
set of parameters in principle cannot be a proof of quality of a theory.
This is no more than an illustration that for the particular set of
parameters inherent (built in) mistakes of the EP theory are numerically small
However, it does not make the EP approach a consistent theory.
In terms of physics, the disregarded lesser and greater projections of self-energy $\Sigma_C$
are responsible for energy exchange between molecular vibrations caused by cavity mode induced
effective interaction. Among other parameters, intensity of radiation field defines strength of the exchange.
Indeed, intensity of radiation field enters $\Sigma_C^{\gtrless}$ via greater/lesser projections of 
Green function $A$, Eq.(\ref{SigmaC}), which in turn is defined by $\Sigma_A^{\gtrless}$, Eq.(\ref{SigmaAB}), 
whose $\Sigma_{rad}$ contribution, Eq.(\ref{Sigma_rad_ph}), is directly proportional to the field intensity.
Thus, it is not surprising that at low intensities the mistake of the EP approach is less pronounced numerically.

\begin{figure}[htbp]
\centering\includegraphics[width=\linewidth]{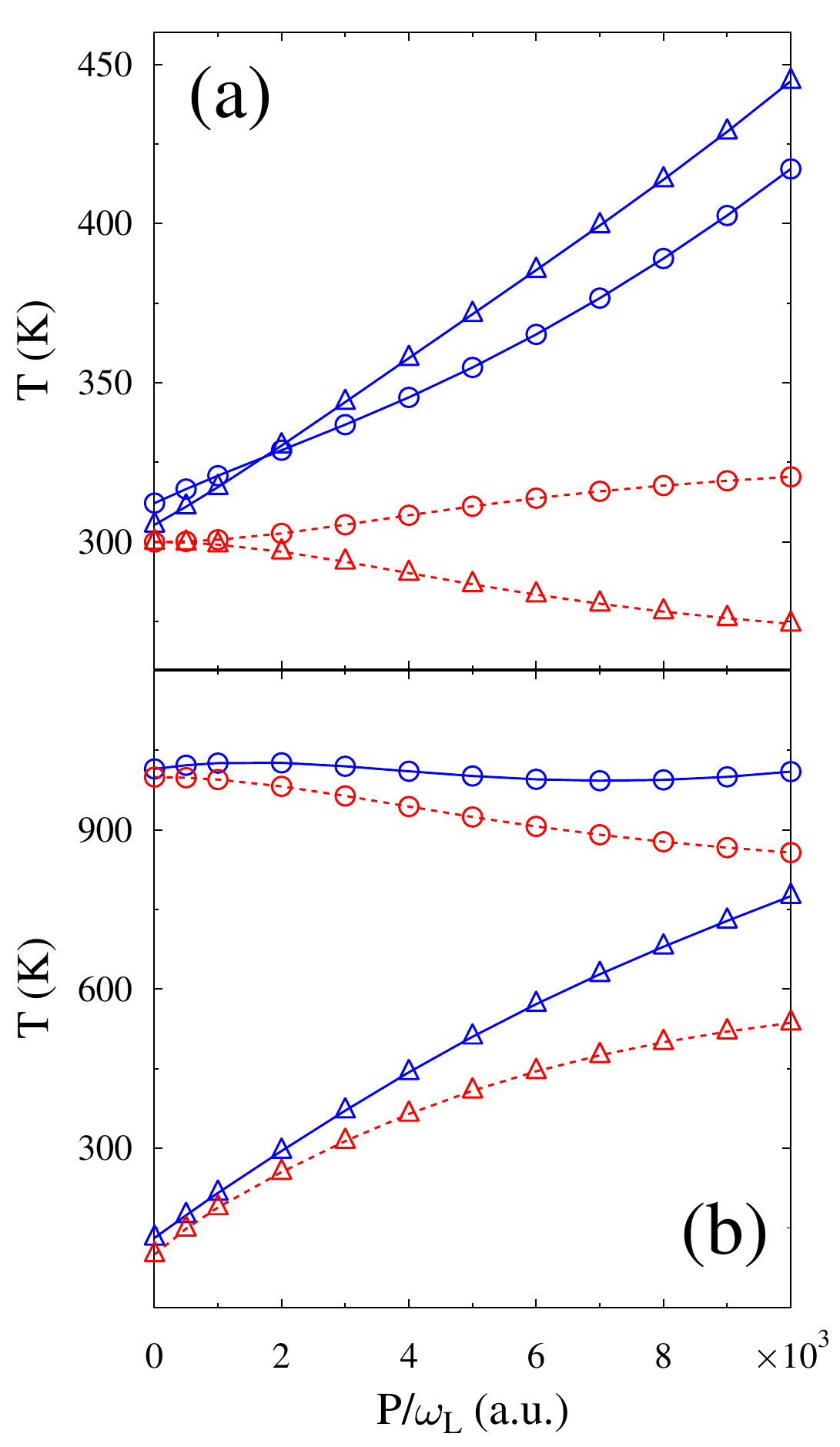}
\caption{\label{fig3}
Effective temperature $T^{eff}$, Eq.(\ref{Teff}), vs the pump laser intensity.
Shown are results of calculations within EP (dashed line, red) and
NEGF (solid line, blue) approaches for $\omega_1$ (circles) and 
$\omega_2$ (triangles). Calculation are performed for (a) $T_1=T_2=300$~K
and (b) $T_1=1000$~K and $T_2=100$~K.
Other parameters are as in Fig.~\ref{fig2}.
}
\end{figure}

Figure~\ref{fig3} shows the effective temperatures of the modes, Eq.(\ref{Teff}), 
as functions of the laser intensity. Discrepancy between EP and NEGF 
in spectrum results naturally leads to quantitative and qualitative differences 
in predicting temperatures of the modes.
For example, Fig.~\ref{fig3}a shows that the EP approach predicts cooling of 
the mode $\omega_2$ when increasing laser frequency, while
proper NEGF simulation show the mode will be heated and its heating
will be more significant than that of mode $\omega_1$.
Similarly, Fig.~\ref{fig3}b shows the EP prediction for monotonic 
cooling of $T_1^{eff}$,
while NEGF yields non-monotonic behavior.

%%%%%%%%%%%%%%%%%%%%%%%%%%%%%%%%%%%%%%%%%%%
\section{Conclusions}\label{conclude}
Starting with the full NEGF treatment 
of two vibrations in a cavity (Fig.~\ref{fig1}) 
we derive an effective non-Hermitian Hamiltonian
formulation and discuss the employed approximations.
The effective Hamiltonian allows to introduce 
exceptional points (degeneracy points in the spectrum of the Hamiltonian).
Together with the input-output formalism the approach is widely used
in the theoretical analysis of optomechanical systems.

The main limitations of the EP approach are related to
\begin{enumerate}
\item Its Markov (delta correlated in time) character, which makes it
inconvenient in treatment of systems with multiple resonances.
\item Its inconsistency in treatment of intra-system interactions
which results in keeping retarded projection of the corresponding self-energy
while disregarding all other projections of the same self-energy. 
\end{enumerate}
For simplicity we disregarded the back action of vibrational modes.
This was done intentionally because incorporation of self-consistency
into the NEGF treatment would make derivation of the EP approach from NEGF
impossible and thus distract from the main message of the paper.
We note that strictly speaking accounting for back action within NEGF is not a matter of choice
but a necessity. Ever since classical works by Kadanoff and Baym in 1960s~\cite{baym_conservation_1961,baym_self-consistent_1962}, 
it is known that  non-self-consistent approximations within NEGF violate conservation laws in the system.
While the limitation of bare perturbation theory is well known in quantum transport, optics community
is less aware about the problem. This limitation was a focus of our previous studies~\cite{gao_optical_2016,mukamel_flux-conserving_2019}.
With the presented derivation we show that the EP approach is not capable to account for 
the self-consistency by its very construction. This is one more limitation of the EP method.
From physics point of view necessity of self-consistent treatment is direct consequence of necessity 
to balance properly energy exchange between the two molecular vibrations. 
Steady-state situation results from multiple energy exchanges.
While the effect is not expected to be very important due to the large difference
in corresponding frequencies, in other systems 
(e.g. in quantum transport applications)
inherent lack of self-consistency in the EP treatment may be one more source
of mistake.

We compare the NEGF and EP predictions for parameters chosen to
represent single-molecule in a cavity.
EP is shown to miss important information which (for the chosen parameters)
leads to qualitatively incorrect predictions of spectra and effective
temperatures of the vibrational modes.
Note that the discrepancy between NEGF and EP results is pronounced
at strong laser fields, where quantum effects (missed by the EP approach)
are not the main factor. 

While our discussion was focused on optomechanics (as a convenient
model to show limitations of the EP treatment)
similar conclusions are relevant for standard polaritonic treatment:
while number of peaks in polaritonic spectrum will be the same 
within the EP and a more rigorous treatments (3 in a similar polaritonic model),
their positions, heights, and linewidths may differ significantly.
Note that similar to situation with quantum transport where 
for non-interacting systems the full NEGF treatment can be substituted with much simpler scattering theory 
based (Landauer-Butiker) approach, also in polaritonic systems with many molecules situation
depends on level of the theory involved. If one considers a set of non-interacting molecules
each individually coupled to cavity mode with mode-induced inter-molecular interaction disregarded 
(that is a non-interacting set of molecules), existing scattering theory based approaches will be capable 
to describe such a situation. However, the moment one is willing to introduce any sort of interaction
into consideration (besides individual molecule mixing with radiation field mode resulting in polariton)
with energy redistribution between the system degrees of freedom 
existing techniques will fail due to the same reasons as considered in the manuscript.
Note also that the conclusions are equally applicable to any order of optical processes in nanocavity systems.
Indeed, the very classification of spectroscopies (processes of particular order) 
accepted in optics community is not applicable to open systems. 
One can introduce a somewhat meaningful analog
of such classification by separating the total photon flux into `different order' contributions
(please, note that such separation is not rigorous due to the same reasons why one cannot rigorously
separate total electronic current into elastic and inelastic fluxes - see e.g. Ref.~\cite{caroli_direct_1972}). 
We discussed the issue in our previous publication~\cite{mukamel_flux-conserving_2019}. 
However, such separation is a post-processing. That is, central object of study in NEGF is the total photon flux,
which includes in it all contributions (optical processes of all orders).

In summary, a careful analysis of the involved approximations
should be performed prior to employing the EP method (and concept of exceptional 
points) in treatment of nanoscale systems.
In particular, for systems with several participating resonances,
systems with significant redistribution of population between 
degrees of freedom (DOF)
or with pronounced effect of back action between the DOFs
EP formalism will not be accurate.

%%%%%%%%%%%%%%%%%%%%%%%%%%%%%%%%%%%%%%%%%%%
\begin{acknowledgments}
This material is based upon work supported by the National Science Foundation
  under Grant No. CHE-2154323 (M.G.) and Grant No. CHE-2246379 (S.M.)
\end{acknowledgments}
%%%%%%%%%%%%%%%%%%%%%%%%%%%%%%%%%%%%%%%%%%%

%%%%%%%%%%%%%%%%%%%%%%%%%%%%%%%%%%%%%%%%%%%
%\bibliography{j_abbrev,ep.bib}
%%%%%%%%%%%%%%%%%%%%%%%%%%%%%%%%%%%%%%%%%%%


\end{document}